\documentclass{ws-procs11x85}
\usepackage{multicol,float} 

\def\araa{ARA\&A}%
\def\apj{ApJ}%
\def\apjl{ApJ}%
\def\aaps{A\&AS}%
\def\mnras{MNRAS}%
\def\nat{Nature}%

\begin{document}

\title{EMISSION FROM LARGE-SCALE JETS IN QUASARS}

\author{YASUNOBU  UCHIYAMA}

\address{Department of High Energy Astrophysics, 
ISAS/JAXA, \\ 3-1-1 Yoshinodai, Sagamihara, Kanagawa, 229-8510, Japan\\ 
E-mail: uchiyama@astro.isas.jaxa.jp}

\begin{abstract}
We consider the emission processes in 
the large-scale jets of powerful quasars 
based on the results obtained with the VLA, {\it Spitzer}, {\it Hubble}, and 
{\it Chandra}. 
We show that two well-known jets, 3C~273 and PKS~1136$-$135, 
 have two distinct spectral components on large-scales: 
(1) the low-energy (LE) synchrotron 
spectrum extending from radio to infrared, and (2) the high-energy (HE)
component arising from optical and extending to X-rays. 
The X-ray emission in quasar jets is often attributed to 
inverse-Compton scattering of cosmic microwave background (CMB) photons 
by radio-emitting electrons in a highly relativistic jet. However, 
recent data prefer 
synchrotron radiation by a second distinct population 
as the origin of the HE component. 
We anticipate that optical polarimetry 
with {\it Hubble} will establish the synchrotron nature of the HE component. 
Gamma-ray observations with \emph{GLAST} (renamed as the \emph{Fermi} 
Gamma-ray Space Telescope), 
as well as future TeV observations, are expected to place 
important constraints on the jet models. 
\end{abstract}

\keywords{galaxies: jets --- 
quasars: individual(3C 273, PKS~1136$-$135) ---
radiation mechanisms: non-thermal }

\bodymatter

\begin{multicols}{2}
\section{Introduction}

The emission processes responsible for the spectral energy distributions (SEDs)
of large-scale quasar jets, constructed 
using \emph{Spitzer}, \emph{HST} and \emph{Chandra} data, 
are the subject of active debate \cite{HK06}. 
The X-ray intensity relative to the radio synchrotron flux 
is generally too high to be explained by synchrotron-self-Compton \cite{Cha00}.
The SED at radio, optical, and X-ray 
traces an inflected shape, which 
disfavors the interpretation of X-rays as due to 
synchrotron radiation from a single population of  electrons.

Inverse-Compton (IC) scattering of CMB photons by
high-energy electrons (with electron Lorentz factor of $\gamma_e \sim 30$) 
in a highly relativistic jet with bulk Lorentz factor $\Gamma \sim 10$ 
initially seemed an attractive 
way to explain the observed X-ray emission \cite{Tav00,CGC01}, 
but this process is also not free of problems \cite{AD04}.
Finally, the X-rays may arise from synchrotron
radiation from extremely energetic protons \cite{Aha02}. 
Determining which of these emission mechanisms produces the
observed X-ray jets in powerful quasars is a strong
motivation for more observations of radio-loud quasars,
and has resulted in a rapid increase 
in the number of known X-ray jets. 

In this work, we highlight new results based on the multiwavelength 
data 
from the \emph{VLA}, \emph{Hubble}, \emph{Spitzer} and \emph{Chandra}, 
aiming at identifying the radiation mechanisms operating in powerful quasar jets.
The jets in the nearest quasar 3C~273 ($z=0.158$) 
and  in the lobe-dominated quasar PKS~1136$-$135
($z=0.554$) are selected for our detailed analysis. 

\section{Quasar Jets in 3C 273 and PKS~1136$-$135}

\subsection{Image}

\begin{figure}[H] 
\centerline{\psfig{file=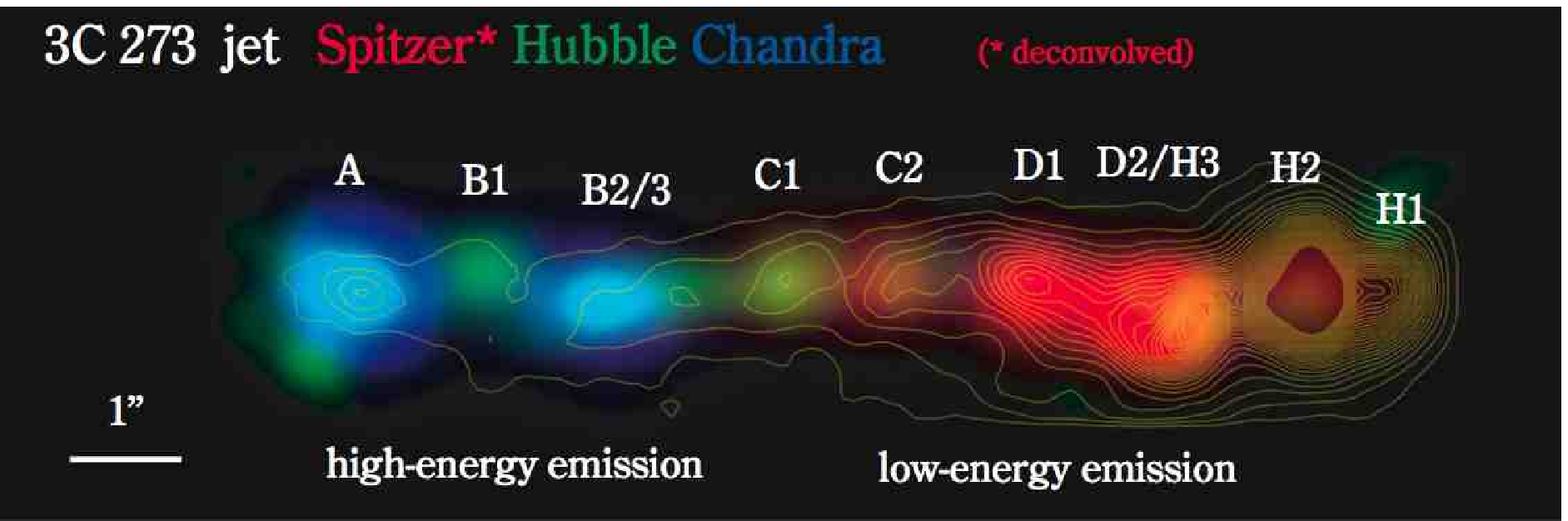,width=8cm}}
\centerline{\psfig{file=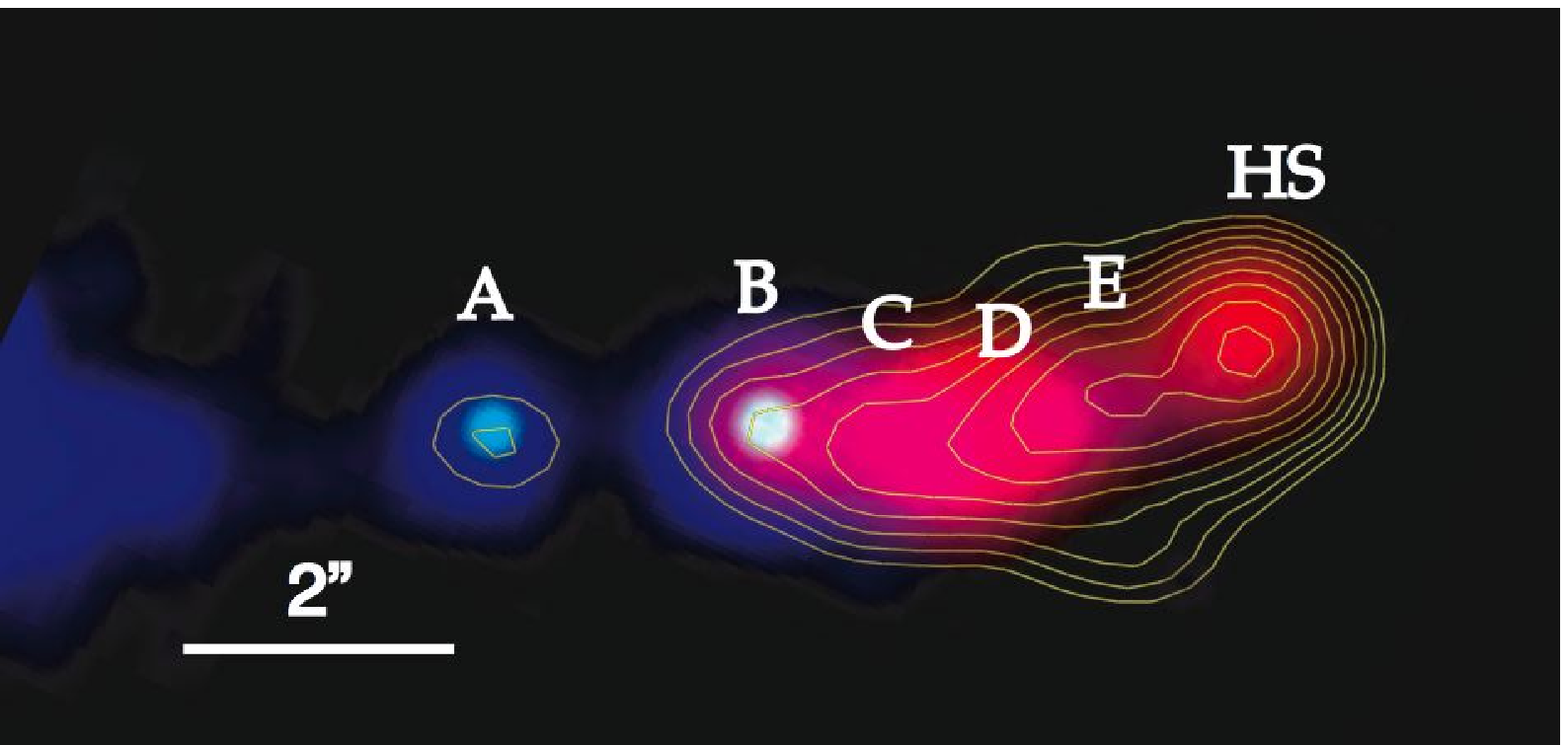,width=8cm}}
\caption{(Top) 
\emph{Spitzer}-\emph{HST}-\emph{Chandra} composite image of
the jet in 3C~273 (Ref.~\refcite{Uch06}):
\emph{Spitzer} ``reconstructed" $3.6\ \mu\rm m$ (\emph{red}), 
\emph{HST} ``UV excess" (\emph{green}), and 
\emph{Chandra} 0.4--6 keV (\emph{blue}). 
The \emph{VLA} 2 cm radio contours are superposed on the image, 
with the strongest radio source H2 being truncated. 
(Bottom) Composite image of the jet in PKS~1136$-$135 
(Ref.~\refcite{Uch07}):
\emph{Spitzer} ({\it red}), \emph{HST} ({\it green dots} 
on knots A and B), and \emph{Chandra} ({\it blue}). 
 The superposed contours are from the \emph{VLA} 8.5 GHz.
Knot A is closest to the quasar core in both cases.}
\label{fig:3c273image}
\end{figure}

Figure~\ref{fig:3c273image} (from Ref.~\refcite{Uch06}) 
presents a three-color image of the large-scale jet in 3C~273 
made with 
\emph{Spitzer}, \emph{HST}, and \emph{Chandra} 
(with the \emph{VLA} contours overlaid). 
The \emph{Spitzer} photometry at $3.6\ \mu\rm m$ is 
illustrated as a series of best-fitted PSFs of every knot 
to restore a resolution similar to the X-rays.
The \emph{HST} image represents  an ``UV excess" map 
emphasizing the near-ultraviolet light.
The inner knots (A--B3) closer to the quasar core are bright in both 
the UV excess and X-rays, 
while the outer knots (C2--H3) are bright in the mid-infrared. 
Recent far UV imaging at 150 nm with \emph{HST} indeed 
confirmed that the inner knots are bright in UV \cite{Jes07}. 

In Fig.~\ref{fig:3c273image} 
we also present a three-color multifrequency image of 
the jet in quasar PKS~1136$-$135 (from Ref.~\refcite{Uch07}). 
At the positions of optically bright knots A and B we depict a dot 
to indicate secure detections with \emph{HST} \cite{Sam06a}.
The brightness patterns along the jet in various bands appear 
to be broadly similar to the  3C~273 jet. 
Like 3C~273, 
the jet knots can be divided up into 
two parts: 
the inner knots (A and B) and the  outer knots (C, D, and E).
The inner knots are bright in both the optical and X-rays, and as such 
they are high-energy dominated.

\begin{figure}[H] 
\centerline{\psfig{file=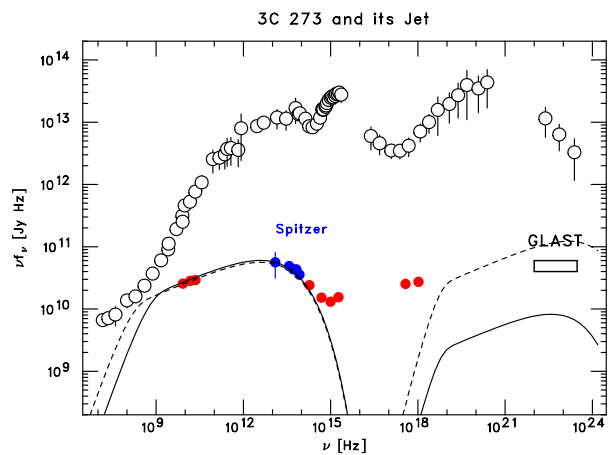,width=8cm}}
\caption{Broadband SEDs of the core and large-scale jet in 3C~273. 
The quasar core (open circles) shows synchrotron and IC components of the 
small-scale jet as well as an accretion disk component in the optical \cite{Tur99}. 
The large-scale jet (accumulated over the entire length from knot A to H1) 
shows two components. 
The curves represent synchrotron radiation in the radio--infrared and 
its IC counterpart in gamma-rays: for 
a jet Doppler factor of $\delta = 4$ (solid curves) and 
$\delta = 8$ (dashed curves). }
\label{fig:3c273SED}
\end{figure}

\subsection{SED}

In Fig.~\ref{fig:3c273SED}, 
the SEDs of both the quasar core and the jet in 3C~273 
are shown. 
The fluxes of the quasar core, originated in a small-scale jet and 
an accretion disk of the central supermassive black hole, are 
described in Ref.~\refcite{Tur99}. 
The jet SED is a ``summed" spectrum for all the knots in Fig.~\ref{fig:3c273image} 
(Refs.~\refcite{Uch06,Jes06,Jes07}).
Here, we added new (unpublished) 
data points at 4.8, 8.0 and 24 $\mu$m in the infrared
 using the \emph{Spitzer} IRAC and MIPS. 
A very wide band from mid IR (24 $\mu$m) to far UV (0.15  $\mu$m) 
is covered with \emph{HST} and \emph{Spitzer}, revealing a  spectral 
character of quasar jets with unprecedented details. 

\begin{figure}[H] 
\centerline{\psfig{file=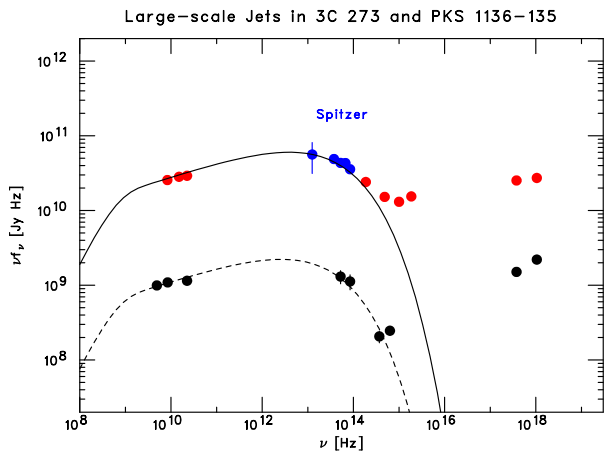,width=8cm}}
\caption{SEDs of the large-scale jets in 3C~273 (taken from Fig.~\ref{fig:3c273SED}) 
and PKS~1136$-$135 (black filled circles: Ref.~\refcite{Uch07}).
The jet in PKS~1136$-$135 (fluxes extracted from knots A to E) 
has a two-component SED similar to 3C~273. 
The curves represent synchrotron radiation models.}
\label{fig:pks1136SED}
\end{figure}

We compare SEDs of the large-scale jets in 3C~273 and PKS~1136$-$135 
in Fig.~\ref{fig:pks1136SED}. The two SEDs share common features. 
The optical and UV fluxes of the jet have a clear excess over the extrapolation 
from the infrared wavelengths, 
indicating that two spectral components cross over at optical wavelengths.
We identify two spectral components as follows: 
(1) the low-energy (LE) synchrotron spectrum extending from radio to infrared 
with a spectral cutoff at $\sim 5\times 10^{13}$ Hz, 
and (2) the high-energy (HE) component arising in the optical and smoothly 
connecting to the X-ray flux. 
The two-component nature of the jet SED becomes more apparent if we 
construct knot-by-knot SEDs (see Refs.~\refcite{Uch06,Jes07}). 
Each SED shows the  two components with varying relative strengths such that 
the low-energy part becomes more  dominant with increasing 
distance from the quasar core. 
The position of the spectral cutoff of the LE component 
seems similar from one knot to another. 
Interestingly, in 3C~273, the spectral slopes of 
the HE component change from $\alpha_{\rm HE} \simeq 0.7$ (inner knots) to 
$\alpha_{\rm HE} \simeq 1.0$ (outer knots). 

\section{Discussion}
\subsection{LE component}

Let us deduce some physical parameters of the LE spectral component, 
which is synchrotron radiation by relativistic electrons in the jet. 
The observed jet radiation is enhanced by  Doppler beaming.
(The Doppler factor is defined as $\delta \equiv [\Gamma(1-\beta\cos\theta)]^{-1}$ 
with $\beta c$ the velocity of the jet, 
 $\Gamma = (1-\beta^2)^{-1/2}$ the bulk Lorentz factor of the jet, and 
$\theta$ the observing angle with respect to the jet direction.)

The LE synchrotron emission has a high-energy cutoff
at $\nu_c \simeq 5\times 10^{13}$ Hz, 
 presumably 
associated with the maximum energy of the electron population.
The maximum energy can be deduced as 
$E_{\rm max} \sim 0.2\ (B_{-4}\delta)^{-0.5}\ \rm  TeV$, 
where  $B_{-4} = B/(0.1\ \rm mG)$ is the comoving magnetic field strength.
The equipartition magnetic field is $B \delta \sim 0.1$ mG for 3C~273 and 
$B \delta \sim 0.7$ mG for PKS~1136$-$135.

By extrapolating VLA radio flux of knot H1/H2 in 3C~273 
to lower frequencies (10--100 MHz), 
we found that a low energy   ``cutoff"  or ``break" should be present as well. 
Otherwise, the jet radio flux around $\sim 100$ MHz  exceeds  even the total 
radio flux of 3C~273. 
It would be reasonable to assume that such a cutoff/break is common to all the knots. 
We adopt $E_{\rm min} \sim 1000 m_ec^2$ in Fig.~\ref{fig:3c273SED} to 
introduce the low energy cutoff. 
Also, the observed synchrotron spectra from the radio to optical frequencies 
are likely to be formed in a  ``cooling regime", 
and therefore a cooling break may appear in low-frequency radio bands.

Using the relation $s_{\rm LE} = 2\alpha_{\rm LE}$, 
where $s_{\rm LE}$ is acceleration index, 
the observed index of $\alpha_{\rm LE} \simeq 0.7$ is translated 
into the electron acceleration index of $s_{\rm LE} \simeq 1.4$, 
close to the hardest possible index  ($s=1.5$) in the nonlinear 
shock acceleration theory.
Here we assumed particle acceleration taking place in each knot,
where relativistic electrons are accumulated over a certain timescale of 
$10^5$ to $10^7$ yr. 
Assuming that the number of radio-emitting electrons is balanced by 
the number of cold protons in the jet, we estimated a jet kinetic power 
of $L_{\rm kin} \sim 5\times 10^{44}\ \rm erg\ s^{-1}$ 
with $\delta = \Gamma = 4$ and the equipartition magnetic field. 

\subsection{HE component: IC/CMB or synchrotron?}

The beamed IC scenario \cite{Tav00,CGC01} was initially considered 
to be an attractive 
 explanation of the X-ray-dominated component of the SED 
for jet knots in radio-loud quasars, like 3C~273 (Ref.~\refcite{Sam01}) 
and PKS~1136$-$135 (Refs.~\refcite{Sam06a,Tav06}).
We argued for 
a direct connection between the optical and X-ray fluxes \cite{Uch06}. 
The same mechanism should 
be responsible for the optical and X-ray emissions.
If one adopts the beamed IC model for the X-rays, 
the energy distribution of electrons has to continue down to very 
low energies of $E_{\rm min} \sim m_e c^2$ to account for the optical emission. 

In the framework of the beamed IC radiation, 
it is interesting to note that 
the jet exhibits 
the evolution of multiwavelength emission along the jet 
as shown in Fig.~\ref{fig:3c273image}. 
The increase of 
radio brightness is accompanied by the decrease of X-ray 
brightness towards downstream, which 
 can be understood in terms of 
deceleration of the jet \cite{GK04,Sam06a,Tav06}.
If the beamed IC scenario is correct, we are able to study 
the flow structure of the large-scale jets. 

New results indicate, however, that the beamed IC scenario faces several  
difficulties. 
First, in the outer knots of the 3C~273 jet, the radio spectral index of 
$\alpha_{\rm LE}\simeq 0.7$  
disagrees with  the HE spectral index of 
$\alpha_{\rm HE}\simeq 1.0$ (see Refs.~\refcite{Uch06,Jes06}).
Second, the parameters required in the beamed IC/CMB model are 
often too extreme. 
In the case of 3C~273, 
we obtained a Doppler factor of $\delta = 30$ for knot A 
and a jet kinetic power of 
$L_{\rm kin} \simeq 1\times 10^{48}\ \rm erg\ s^{-1}$. 
The obtained Doppler factor seems too large for 3C~273, 
since relatively large contributions from 
the un-beamed components are observed in the core emission  
(see Fig.~\ref{fig:3c273SED}) and 
accretion-disk emission appears to be comparable to the jet emission
in the X-ray regime \cite{GP04}.  
In the case of  PKS~1136$-$135, 
we obtained a Doppler factor of $\delta = 19$, which is also 
difficult to reconcile with the lobe-dominated nature of the quasar. 

Instead of the beamed IC, 
synchrotron radiation by a second population of high-energy electrons 
may account for the HE component. 
Then the spectral index of the HE component, 
$\alpha_{\rm HE} \simeq 0.5\mbox{--}1.0$ in 3C~273 and PKS~1136$-$135, 
corresponds to $s_{\rm HE} \simeq 1.0\mbox{--}2.0$. 
However, $s_{\rm HE} \simeq 1.0$ would be incompatible with 
diffusive shock acceleration theory. 
This brings into question the idea that the second electron population 
is accelerated through diffusive shock acceleration.
The second synchrotron component 
may instead be due to turbulent acceleration operating in the shear layers \cite{SO02}. 
Unlike the shock acceleration, 
a hard spectrum, $s < 1.5$, can be expected 
to form in the case of turbulent acceleration (i.e., second-order 
Fermi acceleration). 
This scenario should be tested by future TeV observations (see below).

\subsection{GLAST (Fermi-LAT) observations}

As detailed in Ref.~\refcite{Geo06}, GLAST measurements of 
the quasar jets will provide us with new  information (or constraints) 
about the jet emission. 
The infrared-emitting electrons in a relativistic  jet inevitably 
 emit GeV $\gamma$-rays through an IC/CMB process (see Fig.~\ref{fig:3c273SED}).
The total radiative 
output is dominated by the multi-GeV $\gamma$-rays and 
their predicted flux with $\delta \sim 7$  
exceeds  the 1-yr sensitivity offered by the GLAST LAT 
in the case of 3C 273.
Also, the observations of  PKS~1136$-$135 
with GLAST may be able to test the IC/CMB hypothesis, since 
the beamed IC model with $\delta \sim 20$ predicts the observable GeV flux 
\cite{Uch07}.
Note that steadiness and hard spectrum can distinguish 
the large-scale emission from the  core emission. 

Moreover, if the HE fluxes are due to synchrotron radiation by 
electrons, the UV/X-ray-emitting electrons in a relativistic  jet 
produce IC-upscattered TeV $\gamma$-rays. 
This offers a potential diagnostic tool to distinguish the origin of the HE component. 

\subsection{HST polarimetry}

\begin{figure}[H] 
\centerline{\psfig{file=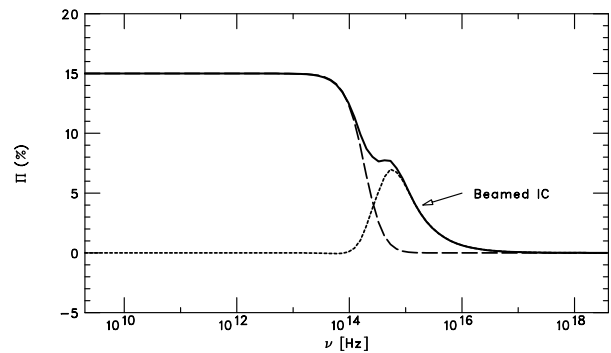,width=8cm}}
\caption{The polarization degree in the synchrotron$+$IC model 
of knot A in the 3C 273 jet. 
The Doppler and Lorentz factor of a jet are $\delta = \Gamma = 15$. 
The minimum electron energy: $E_{\rm min} = 2m_ec^2$.  } 
\label{fig:pol}
\end{figure}

Finally we emphasize  that optical polarimetry can be an effective 
 way of discriminating the radiation models  
responsible for the optical-to-X-ray emission of  the jets.
In the beamed IC interpretation, 
the optical and X-ray emission are due to Compton up-scattering 
off the amplified CMB by high-energy electrons of 
$\gamma_{e} \sim 3$ (optical) and $\gamma_{e} \sim 100$ (X-ray).
Unlike in the case of synchrotron models, 
the X-rays are expected to be {\it unpolarized} and 
the optical light is nearly unpolarized at most a few percent of polarization 
(see Fig.~\ref{fig:pol}).
Precise polarization measurements in the optical can in principle verify (or discard) the 
beamed IC model. Unfortunately, there have been no useful polarization 
observations of quasar jets with {\it HST} so far. 
Only for 3C 273, early {\it HST} polarimetry of the jet was done but 
with the pre-COSTAR FOC and low significances \cite{Tho93}.
Recently, \emph{HST} polarimetry of PKS~1136$-$135 has been approved 
(PI: Eric Perlman). 
New polarimetry of quasar jets on large-scales will provide  important 
clues as to the origin of the HE component. 

\end{multicols}

\begin{thebibliography}{9}


\bibitem{HK06} Harris, D. E., \& Krawczynski, H. 2006, \araa, 44, 463
\bibitem{Cha00} Chartas, G., et al.\  2000, \apj, 542, 655
\bibitem{Tav00} Tavecchio, F., 
Maraschi, L., Sambruna, R. M., Urry, C. M. 
2000, \apjl, 544, L23
\bibitem{CGC01} Celotti, A., Ghisellini, G., \& Chiaberge, M.
2001, \mnras, 321, L1 
\bibitem{AD04} Atoyan, A. M., \& Dermer, C. D. 2004, \apj, 613, 151
\bibitem{Aha02} Aharonian, F. A. 
2002, \mnras, 332, 215 
\bibitem{Uch06} Uchiyama, Y., et al.\  2006, \apj, 648, 910
\bibitem{Uch07} Uchiyama, Y., et al.\ 2007, \apj, 661, 719 
\bibitem{Jes07} Jester, S., et al.\ 2007, \mnras, 380, 828 
\bibitem{Sam06a} Sambruna, R. M., et al.\  2006, \apj, 641, 717
\bibitem{Tur99} T{\"u}rler, M., et al.\ 1999, \aaps, 134, 89
\bibitem{Jes06} Jester, S., Harris, 
D.~E., Marshall, H.~L., \& Meisenheimer, K.\ 2006, \apj, 648, 900 
\bibitem{Sam01} Sambruna, R. M., et al.\ 2001, \apj, 549, L161
\bibitem{Tav06} Tavecchio, F., et al.\  2006, \apj, 641, 732
\bibitem{GK04} Georganopoulos, M., \& Kazanas, D. 2004, \apjl, 604, L81
\bibitem{GP04} Grandi, P., \& Palumbo, G. G. C.
2004,  Science, 306, 998
\bibitem{SO02} Stawarz, \L., \& Ostrowski, M. 2002, \apj, 578, 763
\bibitem{Geo06} Georganopoulos, M.,
Perlman, E., Kazanas, D., \& McEnery, J.\
2006, \apjl, 653, 5 
\bibitem{Tho93} Thomson, R.~C., Mackay, 
C.~D., \& Wright, A.~E.\ 1993, \nat, 365, 133 


\end{thebibliography}
\end{document}